\documentclass[journal,11pt,draftclsnofoot,onecolumn]{IEEEtran}
\usepackage{latexsym}
\usepackage{amsfonts}
\usepackage{graphicx} 
\usepackage{subfigure}

\usepackage{amssymb}
\usepackage{stmaryrd}
\usepackage{natbib}
\usepackage{enumerate}
\usepackage{amsmath}

\newtheorem{Teorema}{\bf Theorem}

\newcommand{\dfn}{\triangleq}
\def\BibTeX{{\rm B\kern-.05em{\sc i\kern-.025em b}\kern-.08em
    T\kfern-.1667em\lower.7ex\hbox{E}\kern-.125emX}}
\begin{document}

\title{Two adaptive rejection sampling schemes for probability density functions log-convex tails}
\author{Luca Martino and Joaqu\'{\i}n M\'{\i}guez\\
Department of Signal Theory and Communications, Universidad Carlos III de Madrid.\\
Avenida de la Universidad 30, 28911 Legan\'es, Madrid, Spain.\\
E-mail: {\tt luca@tsc.uc3m.es, joaquin.miguez@uc3m.es}}
\maketitle
\IEEEpeerreviewmaketitle

\begin{abstract}
Monte Carlo methods are often necessary for the implementation of optimal Bayesian estimators. A fundamental technique that can be used to
generate samples from virtually any target probability distribution is the so-called rejection sampling method, which generates candidate samples from a
proposal distribution and then accepts them or not by testing the ratio of the target and proposal densities. The class of adaptive rejection sampling (ARS) algorithms is particularly interesting because they can achieve high acceptance rates. However, the standard ARS method can only be used with log-concave target densities. For this reason, many generalizations have been proposed.  

In this work, we investigate two different adaptive schemes that can be used to draw exactly from a large family of univariate probability density functions (pdf's), not necessarily log-concave, possibly multimodal and with tails of arbitrary concavity. These techniques are adaptive in the sense that every time a candidate sample is rejected, the acceptance rate is improved. The two proposed algorithms can work properly when the target pdf is multimodal, with first and second derivatives analytically intractable, and when the tails are log-convex in a infinite domain. Therefore, they can be applied in a number of scenarios in which the other generalizations of the standard ARS fail. Two illustrative numerical examples are shown.
\end{abstract}

\section{Introduction}
Monte Carlo methods are often necessary for the implementation of optimal Bayesian estimators and, several families of techniques have been proposed \cite
{Doucet01b,Gilks95bo,Kunsch05,Ahrens95} %\cite{Chopin09,Doucet01b,Gilks95bo,Kunsch05,Philippe03}, %including the Markov Chain Monte Carlo (MCMC) \cite{Fitzgerald01}[3, 7, 15] and particle filtering \citep{Djuric03,Kunsch05} families of algorithms,  
that enjoy numerous applications.
A fundamental technique that can be used to generate independent and identically distributed (i.i.d.) samples from virtually any target probability distribution is the so-called rejection sampling method, which generates candidate samples from a proposal distribution and then accepts them or not by testing the ratio of the target and proposal densities. 

Several computationally efficiencient methods have been designed in which samples from a scalar random variable (r.v.) are accepted with high probability.
%we wish to design methods in which samples are accepted with high probability. 
The class of adaptive rejection sampling (ARS) algorithms \cite
{Gilks92} is particularly interesting because high acceptance rates can be achieved. The standard ARS algorithm yields a sequence of proposal functions that actually converge towards the target probability density distribution (pdf) when the procedure is iterated. As the proposal density becomes closer to the target pdf, the proportion of accepted samples grows (and, in the limit, can also converge to 1). However, this algorithm can only be used with log-concave target densities. A variation of the standard procedure, termed adaptive rejection Metropolis sampling (ARMS) \cite
{Gilks95} can also be used with multimodal pdf's. However, the ARMS algorithm is based on the Metropolis-Hastings algorithm, so the resulting samples form a Markov chain. As a consequence, they are correlated and, for certain multimodal densities, the chain can be easily trapped in a single mode (see, e.g., \cite{MartinoSigPro10}).

Another extension of the standard ARS technique has been proposed in \cite{Hoermann95}, where the same method of  \cite{Gilks92} is applied to $T$-concave densities, with $T$ being a monotonically increasing transformation, not necessarily the logarithm.  However, in practice it is hard to find useful transformations other than the logarithm and the technique cannot be applied to multimodal densities either. The method in \cite{Evans98} generalizes the technique of \cite{Hoermann95} to multimodal distributions. It involves the decomposition of the $T$-transformed density into pieces which are either convex and concave on disjoint intervals and can be handled separately. Unfortunately, this decomposition requires the ability to find all the inflection points of the $T$-transformed density, which can be something hard to do even for relatively simple practical problems. 

More recently, it has been proposed to handle multimodal distributions by decomposing the log-density into a sum of concave and convex functions \cite{Gorur08rev}. Then, every concave/convex element is handled using a method similar to the ARS procedure. A drawback of this technique is the need to decompose the target log-pdf into concave and convex components. Although this can be natural for some examples, it can also be very though for some practical problems (in general, identifying the convex and concave parts of the log-density may require the computation of all its inflection points). Moreover, the application of this technique to distributions with an infinite support requires that the tails of the log-densities be strictly concave.  
Other adaptive schemes that extend the standard ARS method for multimodal densities have been very recently introduced in  \cite
{MartinoStatCo10,MartinoSigPro10}. These techniques include the original ARS algorithm as a particular case and they can be relatively simpler than the approaches in   \cite
{Evans98} or \cite{Gorur08rev} because they do not require the computation of the inflection points of the entire (transformed) target density.
However, the method in \cite{MartinoSigPro10} and the basic approach in \cite{MartinoStatCo10} can also break down when the tails of the target pdf are log-convex, the same as the techniques in \cite{Gilks92,Gorur08rev,Hoermann95}. %Moreover, in \cite{Esperanza10} it is suggested an extension of the standard algorithm to overcome this limitation and in this work,  we take this suggestion and elaborate it to provide a full description of the resulting algorithm.

In this paper, we introduce two different ARS schemes that can be used to draw exactly from a large family of univariate pdf's, not necessarily log-concave and including cases in which the pdf has log-convex tails in an infinite domain. Therefore, the new methods can be applied to problems where the algorithms of, e.g., \cite{Gilks92,Hoermann95,Gorur08rev, MartinoStatCo10} are invalid.

The first adaptive scheme described in this work was briefly suggested in \cite{MartinoStatCo10} as alternative strategy. Here, we take this suggestion and elaborate it to provide a full description of the resulting algorithm. This procedure can be easy to implement and provide good performance as shown in a numerical example. However, it presents some technical requirements that can prevent its use with certain densities, as we also illustrate in a second example.

The second approach introduced here is more general and it is based on the ratio of uniforms (RoU) technique \cite{Devroye86,Kinderman77,Wakefield91}. The RoU method enables us to obtain a two dimensional region $\mathcal{A}$ such that drawing from the univariate target density is equivalent to drawing uniformly from $\mathcal{A}$.  When the tails of the target density decay as $1/x^2$ (or faster), the region $\mathcal{A}$ is bounded and, in such case, we introduce an adaptive technique that generates a collection of non-overlapping triangular regions that cover $\mathcal{A}$ completely. Then, we can efficiently use rejection sampling to draw uniformly from $\mathcal{A}$ (by first drawing uniformly from the triangles). Let us remark that a basic adaptive rejection sampling scheme based on the RoU technique was already introduced in \cite{Leydold00,Leydold03} but it only works when the region $\mathcal{A}$ is strictly convex\footnote{It can be seen as a RoU-based counterpart of the original ARS algorithm in \cite{Gilks92}, that requires the log-density to be concave.}. 
%to sample uniformly from $\mathcal{A}$ when $\mathcal{A}$ is a strictly concave set. 
The adaptive scheme that we introduce can also be used with non-convex sets. 

%The proposed techniques are very flexible, can be applied to a broad class of densities (possibly multimodal) and yield independent and identically distributed (i.i.d.) samples coming exactly from the desired probability distribution.  Moreover, the techniques are adaptive in the sense that every time a candidate sample is rejected, the approximation of the region $\mathcal{A}$ can be improved using the rejected point. In this way, we can achieve a convenient trade-off between the sampling efficiency  (that increases with the accuracy of the approximation) and the computational cost (that also grows as the accuracy is improved). 

The rest of the paper is organized as follows. The necessary
background material is presented in Section \ref{secBack}. The first adaptive procedure is described in Section \ref{AltStrat} while the adaptive RoU scheme is introduced in Section \ref{SecAdScheme}. In Section \ref{sExample} we present two illustrative examples and we conclude with a brief summary and conclusions in Section \ref{sConclusions}.   

\section{Background}
\label{secBack}
In this Section we recall some background material needed for the remaining of the paper. 
First, in Sections \ref{secRS} and \ref{GARSsect1} we briefly review the rejection sampling method and its adaptive implementation, respectively.  
The difficulty of handling target densities with log-convex tails is discussed in Section \ref{SectCotaPro}.
 Finally, we present the ratio of uniforms (RoU) method in Section \ref{sRoU}.

\subsection{Rejection sampling}
\label{secRS}
Rejection sampling \cite
[Chapter 2]{Devroye86} is a universal method for drawing independent samples from a target density $p_o(x)\geq 0$ known up to a proportionality constant (hence, we can evaluate $p(x) \propto p_o(x)$). Let $\exp\{-W(x)\}$ be an overbounding function for $p(x)$, i.e., $\exp\{-W(x)\}\geq p(x)$. We can generate $N$ i.i.d. samples from $p_o(x)$ according to the standard rejection sampling algorithm:
\begin{enumerate}
\item Set $i=1$.
\item Draw samples $x'$ from $\pi(x)\propto \exp\{-W(x)\}$ and $u'$ from $\mathcal{U}(0,1)$, where $\mathcal{U}(0,1)$ is the uniform pdf in $[0,1]$.
\item If $\frac{p(x')}{\exp\{-W(x)\}}\geq u'$ 
then $x^{(i)}=x'$ and set $i=i+1$, else discard $x'$ and go back to step 2.
\item If $i>N$ then stop, else go back to step 2.  
\end{enumerate}
The fundamental figure of merit of a rejection sampler is the mean acceptance rate, i.e., the expected number of accepted samples over the total number of proposed candidates. In practice, finding a tight overbounding function is crucial for the performance of a rejection sampling algorithm. In order to improve the mean acceptance rate many adaptive variants have been proposed \cite
{Evans98,Gilks92,Gorur08rev,Hoermann95,MartinoStatCo10}.

\subsection{An adaptive rejection sampling scheme}
\label{GARSsect1}

We describe the method in \cite
{MartinoStatCo10}, that contains the the standard ARS algorithm of \cite
{Gilks92} as a particular case. 
Let us write the target pdf as $p_o(x)\propto p(x)=\exp\left\{-V(x;\textbf{g})\right\}$, $\mathcal{D} \subset \mathbb{R}$, where $\mathcal{D} \subset \mathbb{R}$ is the support if $p_o(x)$ and $V(x;\textbf{g})$ is termed the potential function of $p_o(x)$. In order to apply the technique of \cite
{MartinoStatCo10}, we assume that the potential function can be expressed  ass the addition of $n$ terms,
\begin{equation}
\label{potentialuca}
	V(x;\textbf{g}) \dfn \sum_{i=1}^{n}\bar{V}_{i}(g_{i}(x)),
\end{equation}
where the functions $\bar{V}_i$, $i=1,...,n$,  referred to as marginal potentials, are convex while every $g_i(x)$, $i=1,...,n$, is either convex or concave. As a consequence, the potential $V(x;\textbf{g})$ is possible non-convex (hence, the standard ARS algorithm cannot be applied) and can have rather complicated forms.

The sampling method is iterative. Assume that, after the $(t-1)$-th iteration, there is available a set of $m_t$ distinct support points, $\mathcal{S}_t = \{ s_1, s_2, \ldots, s_{m_t} \} \subset \mathcal{D}$, sorted in ascending order, i.e., $s_1 < s_2 < \ldots < s_{m_t}$. From this set, we define $m_t+1$ intervals of the form $\mathcal{I}_0 = (-\infty,s_1]$, $\mathcal{I}_k=[s_k,s_{k+1}]$, $k=1,\ldots,m_t-1$, and $\mathcal{I}_{m_t}=[s_{m_t},+\infty)$. For each interval $\mathcal{I}_k$, it is possible to construct a vector of linear functions $\textbf{r}_k(x) = [r_{1,k}(x),\ldots, r_{n,k}(x)]$ such that $\bar{V}_{i}(r_{i,k}(x)) \leq \bar{V}_{i}(g_{i}(x))$ (see \cite
{MartinoStatCo10}) and, as a consequence, 
\begin{equation}
V(x;\textbf{r}_k) \leq V(x;\textbf{g}), \quad \mbox{for all } x \in {\mathcal I}_k 
\end{equation}
and $k=0,...,m_t$. Hence, it is possible to obtain $\exp\{-V(x;\textbf{r}_k)\} \ge p(x)$, $\forall x\in \mathcal{I}_k$. Moreover, the modified potential $V(x;\textbf{r}_k)$ is strictly convex in the interval ${\mathcal I}_k$ and, therefore, it is straightforward to build a linear function $w_k(x)$, tangent to $V(x;\textbf{r}_k)$, such that $w_k(x) \le V(x;\textbf{r}_k)$ and $\exp\{ -w_k(x) \} \ge p(x)$ for all $x \in \mathcal{I}_k$. 

With these ingredients, we can outline the generalized ARS algorithm as follows.  
\begin{enumerate}
\item \textbf{Initialization.} Set $i=1$, $t=0$ and choose $m_1$ support points, $\mathcal{S}_{1}=\{s_{1},..., \ s_{m_1}\}$.
\item \textbf{Iteration.} For $t \ge 1$, take the following steps.
	\begin{itemize}
	\item From $\mathcal{S}_t$, determine the intervals $\mathcal{I}_0, \ldots, \mathcal{I}_{m_t}$.
	\item For $k=0, \ldots, m_t$, construct $\textbf{r}_k$, $V(x;\textbf{r}_k)$ and $w_k(x)$. 
	\item Let $W_t(x) = w_k(x)$, if $x \in \mathcal{I}_k$, $k \in \{ 0,\ldots,m_t \}$.
	\item Build the proposal pdf $\pi_t(x) \propto \exp\{-W_t(x)\}$.
	\item Draw $x'$ from the proposal $\pi_t(x)$ and $u'$ from $\mathcal{U}(0,1)$.
	\item  If $u' \leq \frac{p(x')}{\exp\{-W_{t}(x')\}}$, then accept $x^{(i)}=x'$ and set $\mathcal{S}_{t+1}=\mathcal{S}_{t}$, $m_{t+1}=m_t$, $i=i+1$.
	\item Otherwise, if $u'> \frac{p(x')}{\exp\{-W_{t}(x')\}}$, then reject $x'$, set $\mathcal{S}_{t+1}=\mathcal{S}_{t}\cup \{x'\}$ and update $m_{t+1}=m_t+1$.
	\item Sort $\mathcal{S}_{t+1}$ in ascending order and increment $t=t+1$. If $i>N$, then stop the iteration.
	\end{itemize}
\end{enumerate}
Note that it is very easy to draw from $\pi_t(x)$ because it consists of pieces of exponential densities. Moreover, every time we reject a sample $x'$, it is incorporated into the set of support points and, as a consequence, the shape of the proposal $\pi_t(x)$ becomes closer to the shape of the target $p_o(x)$ and the acceptance rate of the candidate samples is improved \cite
{MartinoStatCo10}.

%%%%%%%%%%%%%%%%%%%%
\subsection{Log-convex tails}
%%%%%%%%%%%%%%%%%%%%
\label{SectCotaPro}

The algorithm of Section \ref{GARSsect1} breaks down when the potential function $V(x;\textbf{g})$ has both an infinite support ($x\in\mathcal{D}=\mathbb{R}$) and concave tails (i.e, the target pdf $p_o(x)$ has log-convex tails). In this case, the function $W_{t}(x)$ becomes constant over an interval of infinite length and we cannot obtain a proper proposal pdf $\pi_t(x)$. To be specific, if $V(x;\textbf{g})$ is concave in the intervals $(-\infty,s_1]$,  $[s_{m_t},+\infty)$ or both, then $w_0(x)$, $w_{m_t}(x)$, or both, are constant and, as a consequence, $\int_{-\infty}^{+\infty} \exp\{-W_t(x)\} dx = +\infty$. 
  
This difficulty with the tails is actually shared by other adaptive rejection sampling techniques in the literature. A theoretical solution to the problem is to find an invertible transformation $G: \mathcal{D}\rightarrow  \mathcal{D}^*$, where $\mathcal{D}^* \subset \mathbb{R}$ is a bounded set \cite
{Devroye86,Hormann94,Marsaglia84,Wallace76}. Then, we can define a random variable $Y=G(x)$ with density $q(y)$, draw samples $y^{(1)}, \ldots, y^{(N)}$ and then convert them into samples $x^{(1)}=G^{-1}(y^{(1)}), \ldots, x^{(N)}=G^{-1}(y^{(N)})$ from the target pdf $p_o(x)$ of the r.v. $X$. However, in practice, it is difficult to find a suitable transformation $G$, since the resulting density $q(y)$ may not have a structure that makes sampling any easier than in the original setting.

A similar, albeit more sophisticated, approach is to use the method of \cite
{Evans98}. In this case, we need to build a partition of the domain $\mathcal{D}$ with disjoint intervals $\mathcal{D}= \cup_{i=1}^m\mathcal{D}_i$ and then apply invertible transformations  $T_i:\mathcal{D}_i \rightarrow \mathbb{R}$, $i=1,...,m$, to the target function $p(x)$. In particular, the intervals $\mathcal{D}_1$ and $\mathcal{D}_m$ contain the tails of $p(x)$ and the method works correctly if the composed functions $(T_1 \circ p)(x)$ and $(T_m \circ p)(x)$ are concave. However, finding adequate $T_1$ and $T_m$ is not necessarily a simple task and, even if they are obtained, applying the algorithm of \cite
{Evans98} requires the ability to compute all the inflection points of the target function $p(x)$.

%%%%%%%%%%%%%%%%%%%%%%%%%%
\subsection{Ratio of uniforms method}
%%%%%%%%%%%%%%%%%%%%%%%%%%
\label{sRoU}
The RoU method \cite
{Chung97,Kinderman77,Wakefield91} is a sampling technique that relies on the following result.

\begin{Teorema}
Let $q(x)\geq 0$ be a pdf known only up to a proportionality constant. If $(u,v)$ is a sample drawn from the uniform distribution on the set 
\begin{equation}
\label{regionAdef}
\mathcal{A}= \Big\{ (v,u)\in \mathbb{R}^2: 0\leq u \leq \sqrt{q(v/u)}\Big\}, 
\end{equation}
then $x=\frac{v}{u}$ is a sample form $q(x)$. 
\end{Teorema}

{\bf Proof}: See \cite
[Theorem 7.1]{Devroye86}.

Therefore, if we are able to draw uniformly from $\mathcal{A}$, then we can also draw from the pdf $q_o(x) \propto q(x)$.  
The cases of practical interest are those in which the region $\mathcal{A}$ is bounded, and $\mathcal{A}$ is bounded if, and only if, both $\sqrt{q(x)}$ and $x\sqrt{q(x)}$ are bounded. Moreover, the function $\sqrt{q(x)}$  is bounded if, and only if, the target density $q_o(x) \propto q(x)$ is bounded and, assuming we have a bounded target function $q(x)$, the function $x\sqrt{q(x)}$ is bounded if, and only if, the tails of $q(x)$ decay as $1/x^2$ or faster. %However,  it exists some generalizations \cite
{Chung97,Wakefield91} defining different region $\mathcal{A}$ that could be used for fatter tails. 
 
\begin{figure*}[htb]
\centering
\centerline{
   % \subfigure[]{\includegraphics[width=5.6cm]{RoU4.pdf}} 
  %  \subfigure[]{\includegraphics[width=5.6cm]{RoUconq.pdf}}  
    \subfigure[]{\includegraphics[width=5.6cm]{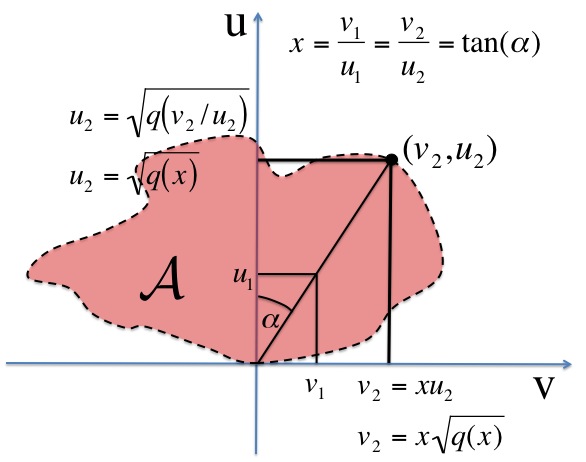}}  
      %  \subfigure[]{\includegraphics[width=5.6cm]{RoU6.pdf}}
  %  \subfigure[]{\includegraphics[width=5.6cm]{RoUconq2.pdf}}  
    \subfigure[]{\includegraphics[width=5.6cm]{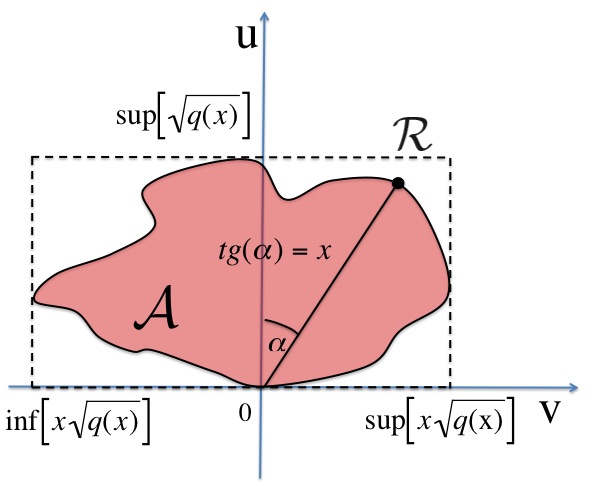}}      
 %  \includegraphics[width=5.6cm]{CotaConvex2.pdf} 	
% \subfigure[]{\includegraphics[width=4.6cm]{TriangleGen.pdf}} 
}
\caption{ \textbf{(a)} A bounded region $\mathcal{A}$ and the straight line $v=xu$ corresponding to the sample $x=\tan(\alpha)$. Every point in the intersection of the line $v=xu$ and the set $\mathcal{A}$ yields the same sample $x$. The point on the boundary, $(v_2,u_2)$, has coordinates $v_2=x\sqrt{q(x)}$ and $u_2=\sqrt{q(x)}$. \textbf{(b)} If the two functions $\sqrt{q(x)}$ and $x\sqrt{q(x)}$ are bounded, the set $\mathcal{A}$ is bounded and embedded in the rectangle $\mathcal{R}$. %\textbf{(c)} A triangular region $\mathcal{A}$ with a vertex at the origin $\textbf{v}_1=(0,0)$ and where the side $\textbf{v}_2-\textbf{v}_3$ has a generic slope. It corresponds to a density of the form $q_o(x)\propto 1/(\lambda x+\beta)^2$ via the RoU method. 
} 
\label{figuras}
\end{figure*}

Figure \ref{figuras} (a) depicts a bounded set $\mathcal{A}$. Note that, for every angle $\alpha\in (-\pi/2,+\pi/2)$ rad, we can draw a straight line that passes through the origin $(0,0)$ and contains  points $(v_i,u_i)\in \mathcal{A}$ such that $x=\frac{v_i}{u_i}=\tan(\alpha)$, i.e., every point $(v_i,u_i)$ in the straight line with angle $\alpha$ yields the same value of $x$.

From the definition of $\mathcal{A}$, we obtain $u_i\leq q(x)$ and $v_i=xu_i \leq x\sqrt{q(x)}$, hence, if we choose the point $(v_2,u_2)$ that lies on the boundary of $\mathcal{A}$, $u_2=\sqrt{q(x)}$ and $v_2=x\sqrt{q(x)}$. Moreover, we can embed the set $\mathcal{A}$ in the rectangular region
\begin{gather}
\begin{split}
\mathcal{R}=\Big\{&(v',u'): 0 \leq u' \leq \sup_x\sqrt{q(x)},  \\
&\inf_x x\sqrt{q(x)}\leq  v' \leq  \sup_x x\sqrt{q(x)} \Big\} ,
\end{split}
\end{gather}
as depicted in Fig. \ref{figuras} (b).

Once $\mathcal{R}$ is constructed, it is straightforward to draw uniformly from $\mathcal{A}$ by rejection sampling: simply draw uniformly from $\mathcal{R}$ and then check whether the candidate point belongs to $\mathcal{A}$.

%In some cases the equation $u=\sqrt{q(v/u)}$ can be solved analytically and the boundary $\mathcal{A}$ can be found explicitly.
%In particular, when $q_o(x)\propto \frac{\tau^2}{(\lambda x+\beta)^2}$  with $x\in [a,b]$, the region $\mathcal{A}$ is a triangle, as depicted in Fig. \ref{figuras} (c), with one vertex at the origin, $\textbf{v}_1=(0,0)$, and the opposite side, $\textbf{v}_2-\textbf{v}_3$, with equation $\lambda v+\beta u=\tau$. 

%%%%%%%%%%%%%%%%%%%%%%%%%%%%%%%%%%%%%%%
\section{Adaptive rejection sampling with log-convex tails}
%%%%%%%%%%%%%%%%%%%%%%%%%%%%%%%%%%%%%%%
\label{AltStrat}

In this section, we investigate a strategy proposed in \cite
{MartinoStatCo10} to obtain an adaptive rejection sampling algorithm that remains valid the tails of the potential function $V(x;{\bf g})$ are concave (i.e, the target pdf $p_o(x)$ has log-convex tails).

Let us assume that for some $j\in \{1,\ldots,n\}$, the pdf defined as  
\begin{equation}
q(x)\propto \exp\{-\bar{V}_j(g_j(x))\}
\end{equation}
is such that: (a) we can integrate $q(x)$ over the intervals $\mathcal{I}_{0},\mathcal{I}_{1},...,\mathcal{I}_{m_t}$ and (b) we can sample from the density $q(x)$ restricted to every $\mathcal{I}_k$.
 To be specific, let us introduce the \textit{reduced} potential
\begin{equation}
V_{-j}(x;\textbf{g})\dfn \sum_{i=1,i\neq j}^{n}\bar{V}_i(g_i(x)),
\end{equation}
attained by removing $\bar{V}_j(g_j(x))$ from $V(x;\textbf{g})$. It is straightforward to obtain lower bounds $\gamma_k\leq V_{-j}(x;\textbf{g})$ by applying the procedure explained in the Appendix to the reduced potential $V_{-j}(x;\textbf{g})$ in every interval $\mathcal{I}_k$, $k=0,...,m_t$. Once these bounds are available, we set $L_k\dfn\exp\{-\gamma_k\}$ and build the piecewise proposal function 
\begin{equation}
\label{propotraStrat}
\pi_t(x) \propto
	\left\{
\begin{array}{l}
 L_0\exp\{-\bar{V}_j(g_j(x))\},  \  \  \forall x\in\mathcal{I}_0, \\
 \vdots \\
L_k\exp\{-\bar{V}_j(g_j(x))\},  \  \  \forall x\in\mathcal{I}_k, \\
 \vdots \\
 L_{m_t}\exp\{-\bar{V}_j(g_j(x))\},  \mbox{ } \forall x\in\mathcal{I}_{m_t}. \\
\end{array}
\right.
\end{equation}
Notice that, for all $x\in \mathcal{I}_k$, we have $L_k\geq \exp\{-V_{-j}(x;\textbf{g})\}$ and multiplying both sides of this inequality by the positive factor $\exp\{-\bar{V}_j(g_j(x))\}\geq 0$, we obtain 
\[L_k\exp\{-\bar{V}_j(g_j(x))\}\geq \exp\{-V(x;\textbf{g})\}, \mbox{ } \mbox{ }\forall x\in\mathcal{I}_k,\]
hence $\pi_t(x)$ is suitable for rejection sampling. 

Finally, note that $\pi_t(x)$ is a mixture of truncated densities with non-overlapping supports. Indeed, let us define the mixture coefficients 
\begin{equation}
\bar{\alpha}_{k}\dfn L_k\int_{\mathcal{I}_k}q(x) dx
\end{equation}
and normalize them as $\alpha_{k}=\bar{\alpha}_{k}/\sum_{k=0}^{m_t}\bar{\alpha}_{k}$. Then,
\begin{equation}
\label{EqPropFun}
\pi_t(x)=\sum_{k=1}^{m_t} \alpha_k q(x) \chi_k(x) 
\end{equation}
where $\chi_k(x)$ is an indicator function ($\chi_k(x)=1$ if $x\in \mathcal{I}_k$ and $\chi_k(x)=0$ if $x\notin \mathcal{I}_k$). The complete algorithm is summarized below. 
\begin{enumerate}
\item \textbf{Initialization.} Set $i=1$, $t=0$ and choose $m_1$ support points, $\mathcal{S}_{1}=\{s_{1},..., \ s_{m_1}\}$. 
\item \textbf{Iteration.} For $t \geq 1$, take the following steps.
	\begin{itemize}
	\item From $\mathcal{S}_t$, determine the intervals $\mathcal{I}_0, \ldots, \mathcal{I}_{m_t}$.
	\item Choose a suitable pdf $q(x)\propto \exp\{-\bar{V}_j(g_j(x))\}$ for some $j\in \{1,...,n\}$. 
	\item For $k=0, \ldots, m_t$, compute the lower bounds $\gamma_k\leq V_{-j}(x,{\bf g})$, $\forall x\in \mathcal{I}_k$, and set $L_k=\exp\{-\gamma_k\}$ (see the Appendix). 
	\item Build the proposal pdf $\pi_t(x) \propto L_k q(x)$ for $x \in \mathcal{I}_k$, $k \in \{ 0,\ldots,m_t \}$.
	\item Draw $x'$ from the proposal $\pi_t(x)$ in Eq. (\ref{EqPropFun}) and $u'$ from $\mathcal{U}(0,1)$.
	\item  If $u' \leq \frac{\exp\{-V_{-j}(x';{\bf g})\}}{L_k}$, then accept $x^{(i)}=x'$ and set $\mathcal{S}_{t+1}=\mathcal{S}_{t}$, $m_{t+1}=m_t$, $i=i+1$.
	\item Otherwise, if $u'> \frac{\exp\{-V_{-j}(x';{\bf g})\}}{L_k}$, then reject $x'$, set $\mathcal{S}_{t+1}=\mathcal{S}_{t}\cup \{x'\}$ and update $m_{t+1}=m_t+1$.
	\item Sort $\mathcal{S}_{t+1}$ in ascending order and increment $t=t+1$. If $i>N$, then stop the iteration.
	\end{itemize}
\end{enumerate}

 When a sample $x'$ drawn from $\pi_t(x)$ is rejected, $x'$ is added to the set of support points $\mathcal{S}_{t+1}\dfn\mathcal{S}_{t}\cup \{x'\}$. Hence, we improve the piecewise constant approximation of $V_{-j}(x;\textbf{g})$ (formed by the upper bounds $L_k$)  and proposal pdf $\pi_{t+1}(x)\propto L_k\exp\{-\bar{V}_j(g_j(x))\}$, $\forall x\in \mathcal{I}_k$, becomes closer to the target pdf $p_o(x)$. 

\begin{figure*}[htb]
\centering
\centerline{
 %   \subfigure[]{\includegraphics[width=6cm]{AdaptiveScheme1fig.pdf}}  
    \subfigure[]{\includegraphics[width=5.5cm]{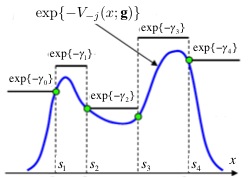}}  
 %     \subfigure[]{\includegraphics[width=6.5cm]{AdaptiveScheme1fig2.pdf}}  
      \subfigure[]{\includegraphics[width=6cm]{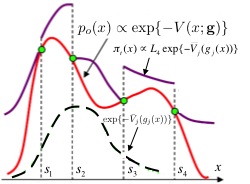}}  
}
\caption{\textbf{(a)} Example of the function $\exp\{-V_{-j}(x;{\bf g})\}$ and its stepwise approximation $L_k=\exp\{-\gamma_k\}$, $k=0,...,m_t=4$, constructed with the proposed technique using four support points $\mathcal{S}_t=\{s_1,s_2,s_3,s_4\}$. \textbf{(b)} Our target pdf $p_o(x)\propto \exp\{-V(x;{\bf g})\}=\exp\{-V_{-j}(x;{\bf g})-\bar{V}_{j}(g_j(x))\}$ obtained by multiplying the previous function $\exp\{-V_{-j}(x;{\bf g})\}$ times $\exp\{-\bar{V}_{j}(g_j(x))\}$ (shown with dashed line). The picture also shows the shape of the proposal density $\pi_t(x)$ consisting of pieces of the function $\exp\{-\bar{V}_{j}(g_j(x))\}$ scaled by the constant values $L_k=\exp\{-\gamma_k\}$. } 
\label{figuras4}
\end{figure*}

Figure \ref{figuras4} (a) illustrates the reduced potential $\exp\{-V_{-j}(x;{\bf g})\}$ and its stepwise approximation $L_k=\exp\{-\gamma_k\}$ built using $m_t=4$ support points. Figure \ref{figuras4} (b) depicts the target pdf $p_o(x)$ jointly with the proposal pdf $\pi_t(x)$, composed by weighted pieces of the function $\exp\{-\bar{V}_j(g_j(x))\}$ (shown, in dashed line). 

This procedure is feasible only if we can find a pair $\bar{V}_j$, $g_j$, for some $j\in\{1,...,n\}$, such that the pdf $q(x)\propto \exp\{-\bar{V}_j(g_j(x))\}$ is 
\begin{itemize}
\item is analytically integrable in every interval $\mathcal{I}\subset \mathcal{D}$, given otherwise the weights $\alpha_1$,...,$\alpha_{m_t}$ in Eq. (\ref{EqPropFun}) cannot be computed in general, and
\item is easy to draw from when truncated into a finite or an infinite interval, since otherwise we cannot generate samples from it.
\end{itemize}

Note that in order to draw from $\pi_t(x)$ we need to be able to draw from and to integrate analytically truncated pieces of $q(x)$.
This procedure is possible only if we have on hand a suitable pdf $q(x)\propto \exp\{-\bar{V}_j(g_j(x))\}$.  

The technique that we introduce in the next section, based on ratio of uniforms method, overcomes these constraints.

%does not need this condition (it is more general, as shown in the numerical example in Section \ref{StocVolEj2}).
 
%%%%%%%%%%%%%%%%%%%%%%%%%%%%%%%
\section{Adaptive RoU scheme}
%%%%%%%%%%%%%%%%%%%%%%%%%%%%%%%
\label{SecAdScheme}
%To draw from $p(x|\textbf{y})$ removing problems with the infinite tails, we can also apply another adaptive scheme using RoU method. 
%If $p(x|\textbf{y})$ is defined in a finite domain, the easiest approach to use RoU method is to find upper bound for $p(x|\textbf{y})$  building a piecewise constant approximation of the posterior pdf. Then, this rectangular approximation of $p(x|\textbf{y})$ can be transformed to a region formed by triangular pieces using RoU approach covered the region $\mathcal{A}$ corresponding to $p(x|\textbf{y})$. We have seen in Section \ref{SampleTri}, that it is straightforward draw samples uniformly in triangular region, therefore we could apply rejection sampling. But, in general, if the target density is defined in a finite domain and we are able to find upper bounds and so a rectangular approximation the best idea is to apply rejection sampling directly, i.e., actually there are not reasons to use RoU method.

The RoU method can be a useful tool to deal with target pdf's $p_o(x)\propto p(x)=\exp\{-V(x;{\bf g})\}$ where $V(x;{\bf g})$ of the form in Eq. (\ref{potentialuca}) in a infinite domain and with tails of arbitrary concavity. Indeed, as long as the tails of $p(x)$ decay as $1/x^2$ (or faster), the region $\mathcal{A}$ defined by the RoU transformation is bounded and, therefore, the problem of drawing samples from the target density becomes equivalent to that of drawing uniformly from a finite region.  

In order to draw uniformly from $\mathcal{A}$ by rejection sampling, we need to build a suitable set $\mathcal{P}\supseteq \mathcal{A}$ for which we are able to generate uniform samples easily. This, in turn, requires knowledge of upper bounds for the functions $\sqrt{p(x)}$ and $x\sqrt{p(x)}$, as described in Section \ref{sRoU}. Indeed, note that to apply the RoU transformation directly to an improper proposal $\pi_t(x)\propto \exp\{W_t(x)\}$, with $\exp\{W_t(x)\}\geq p(x)$ and $\int_{-\infty}^{+\infty} \pi_t(x)dx\rightarrow +\infty$, provides us an unbounded region $\mathcal{A}$ so that this strategy is useless.

In the sequel, we describe how to adaptively build a polygonal sets $\mathcal{P}_t$ ($t=1,2,....$) composed by non-overlapping triangular subsets that embed the region $\mathcal{A}$. To draw uniformly from $\mathcal{P}_t$, we randomly select one triangle (with probability proportional to its area) and then generate uniformly a sample point from it using the algorithm in the Appendix.
If the sample belongs to $\mathcal{A}$, it is accepted, and otherwise is rejected and the region $\mathcal{P}_t$ is improved by adding another triangular piece.  

In Section \ref{SecondAdaptivesect} below, we provide a detailed description of the proposed adaptive technique. Details of the computation of some bounds that are necessary for the algorithm are given in Section \ref{otherBound} and in the Appendix.

%%%%%%%%%%%%%%%%%%%%%%%%%%%%
\subsection{Adaptive Algorithm}
\label{SecondAdaptivesect}
%%%%%%%%%%%%%%%%%%%%%%%%%%%%

%How explained above and in Section \ref{secRS} in order to draw from the target density $p(x|\textbf{y})$ applying the ratio-of-uniforms method, in general we need to find bounds for the functions $\sqrt{p(x|\textbf{y})}$ and $x\sqrt{p(x|\textbf{y})}$. Specifically, since $p(x|\textbf{y})\geq 0$, we only need the upper bounds  $L^{(1)}\geq \sqrt{p(x|\textbf{y})}$, $L^{(2)}\geq x\sqrt{p(x|\textbf{y})}$  and $L^{(3)}\geq -x\sqrt{p(x|\textbf{y})}$ (note that $-L^{(3)}$ is a lower bound for $x\sqrt{p(x|\textbf{y})}$).

Given a set of support points
\[
\mathcal{S}_{t}=\{s_1,\ldots,s_{m_{t}}\},	
\]
we assume that there always exists $k'\in\{0,...,m_t\}$ such that $s_{k'}=0$, i.e., the point zero is included in the set of support points $\mathcal{S}_{t}$.  Hence, in an interval $\mathcal{I}_k=[s_{k},s_{k+1}]$ the points $s_{k}$ and $s_{k+1}$ are both non-positive or both non-negative. 

We also assume that we are able to compute upper bounds $L^{(1)}_k\geq \sqrt{p(x)}$, $L^{(2)}_k\geq x\sqrt{p(x)}$ (for $x>0$) and $L^{(3)}_k\geq -x\sqrt{p(x)}$ (for $x<0$) within every interval $x\in \mathcal{I}_k=[s_k,s_{k+1}]$, $k=0,...,m_t$, where $\mathcal{I}_0=(-\infty,s_{1}]$ and $\mathcal{I}_{m_t}=[s_{m_t},+\infty)$.

\subsubsection{Construction of $\mathcal{P}_t\supseteq \mathcal{A}$}
\label{secConT}
\begin{figure*}[htb]
\centering
\centerline{
   %\includegraphics[width=7cm]{Triangle.pdf}   
   %\subfigure[]{\includegraphics[width=7cm]{TriangleRoU2.pdf}} 
   \subfigure[]{\includegraphics[width=7cm]{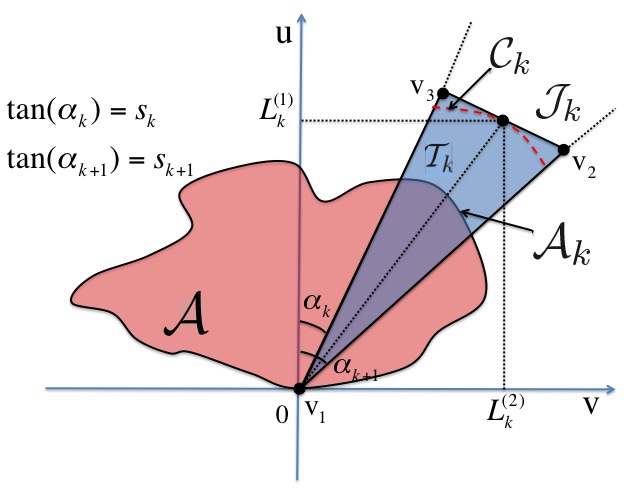}}  
  %   \subfigure[]{\includegraphics[width=7cm]{TriangTodo2.png}}   
 %  \subfigure[]{\includegraphics[width=7cm]{TriangTodo2.pdf}}   
   \subfigure[]{\includegraphics[width=7cm]{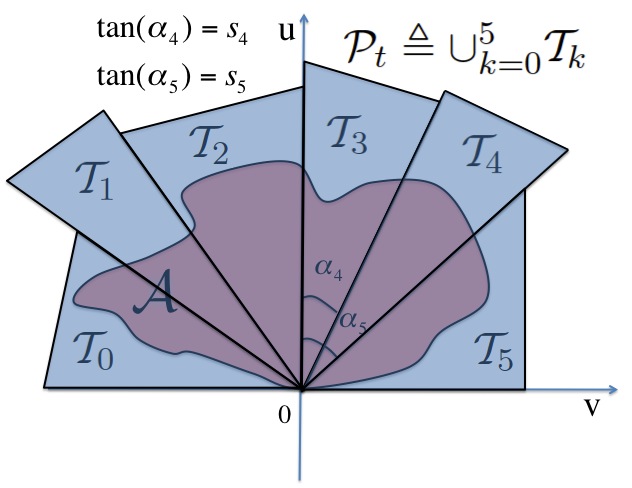}}      	
}
\caption{\textbf{(a)} A region $\mathcal{A}$ constructed by the RoU method and a triangular region $\mathcal{T}_k$ defining by the vertices $\textbf{v}_1$, $\textbf{v}_2$ and $\textbf{v}_3$ built using the upper bounds $L^{(1)}_k$, $L^{(2)}_k$ for the functions $\sqrt{p(x)}$ and $x\sqrt{p(x)}$ for all $x\in \mathcal{I}_k=[s_k,s_{k+1}]$. The dashed line depicts the piece of circumference $\mathcal{C}_k$ with radius $r_k=\sqrt{(L^{(1)}_k)^2+(L^{(2)}_k)^2}$.  The set $\mathcal{T}_k$ embeds the subset $\mathcal{A}_k=\mathcal{A}\cap\mathcal{J}_k$ where $\mathcal{J}_k$  is the cone defined as $\textbf{v}\in\mathcal{J}_k$ if and only if $\textbf{v}=\theta_1 \textbf{v}_2+\theta_2 \textbf{v}_3$ and $\theta_1,\theta_2\geq 0$. \textbf{(b)} Construction of the polygonal region $\mathcal{P}_t= \cup_{k=0}^{5} \mathcal{T}_k$ using $m_t=5$ support points, i.e., $\mathcal{S}_t=\{s_1,s_2,s_3=0,s_4,s_5\}$. Observe that each triangle $\mathcal{T}_k$ has a vertex at $(0,0)$. The set $\mathcal{P}_t$ covers completely the region $\mathcal{A}$ obtained by the RoU method, i.e.,  $\mathcal{A}\subset \mathcal{P}_t$.  
} 
\label{figura2}
\end{figure*}

Consider the construction in Figure \ref{figura2} (a). For a pair of angles $\alpha_k\dfn \arctan(s_k)$ and $\alpha_{k+1}\dfn \arctan(s_{k+1})$, we define the subset $\mathcal{A}_k\dfn \mathcal{A}\cap\mathcal{J}_k$, where $\mathcal{J}_k$ is the cone with vertex at the origin $(0,0)$ and delimited by the two straight lines that form angles $\alpha_{k}$ and $\alpha_{k+1}$ w.r.t. the $u$ axis. Note that, clearly, $\mathcal{A}=\cup_{k=0}^{m_t}\mathcal{A}_k$.

%Let $\alpha_k\dfn \arctan(s_k)$ and $\alpha_{k+1}\dfn \arctan(s_{k+1})$. We define the subset $\mathcal{A}_k$ as the intersection between the region $\mathcal{A}$ and the conic set $\mathcal{J}_k$ composed by the two straight lines forming the two angles $\alpha_k$, $\alpha_{k+1}$ respect to the axis $u$ and the origin of axes $(0,0)$ (in formula, $\mathcal{A}_k\dfn \mathcal{A}\cap\mathcal{J}_k$). Clearly, by definition, we have $\mathcal{A}=\cup_{k=0}^{m_t}\mathcal{A}_k$.

For each $k=0,...,m_t$ the subset $\mathcal{A}_k$ is contained in a piece of circle $\mathcal{C}_k$  ($\mathcal{A}_k\subseteq \mathcal{C}_k$) delimited by the angles $\alpha_k$, $\alpha_{k+1}$ and radius 
\begin{gather}
r_k=\left\{
\begin{split}
\sqrt{\Big(L^{(1)}_k\Big)^2+\Big(L^{(2)}_k\Big)^2}   \ , \  \   \mbox{ }\mbox{ } \mbox{ }\mbox{ }\mbox{ }\mbox{if}\mbox{ }\mbox{ }  \ s_{k},s_{k+1}\geq 0 \\
\sqrt{\Big(L^{(1)}_k\Big)^2+\Big(L^{(3)}_k\Big)^2}  \ , \  \    \mbox{ }\mbox{ }\mbox{ }\mbox{ }\mbox{ }\mbox{if}\mbox{ }\mbox{ }  \ s_{k},s_{k+1}\leq 0 \\
\end{split}
\right.
\end{gather}
also shown (with a dashed line) in Fig. \ref{figura2} (a).
%\[r_k=\sqrt{\Big(L^{(1)}_k\Big)^2+\Big(L^{(2)}_k\Big)^2}\]
%if $s_{k},s_{k+1}\geq 0$ or 
%\[r_k=\sqrt{\Big(L^{(1)}_k\Big)^2+\Big(L^{(3)}_k\Big)^2}\] if $s_{k},s_{k+1}\leq0$, for each $k=0,...,m_t$. 

Unfortunately, it is not straightforward to generate samples uniformly from $\mathcal{C}_k$, but we can easy draw samples uniformly from a triangle in the plane $\mathbb{R}^2$, as explained in the Appendix. Hence, we can choose an arbitrarily a point in the arc of circumference that delimits $\mathcal{C}_k$ (e.g., the point $(L^{(2)}_k,L^{(1)}_k)$ in Fig. \ref{figura2} (a)) and calculate the tangent line to the arc at this point. In this way, we build a triangular region $\mathcal{T}_k$ such that $\mathcal{T}_k \supseteq \mathcal{C}_k  \supseteq \mathcal{A}_k$, with a vertex at $(0,0)$. 

We can repeat the procedure for every $k=0,...,m_t$, with different angles $\alpha_k= \arctan(s_k)$ and $\alpha_{k+1}= \arctan(s_{k+1})$, and define the polygonal region $\mathcal{P}_t\dfn \cup_{k=0}^{m_t} \mathcal{T}_k$ composed by non-overlapping triangular subsets. Note that, by construction, $\mathcal{P}_t$ embeds the entire region $\mathcal{A}$, i.e., $\mathcal{A}\subseteq \mathcal{P}_t$. 

Figure \ref{figura2} summarizes the procedure to build the set $\mathcal{P}_t$. In Figure \ref{figura2}  (a) we show the construction of a triangle $\mathcal{T}_k$ within the angles $\alpha_k=\arctan(s_k)$, $\alpha_{k+1}=\arctan(s_{k+1})$, using the upper bounds $L^{(1)}_k$ and $L^{(2)}_k$ for a single interval $x\in \mathcal{I}_k=[s_k,s_{k+1}]$. Figure \ref{figura2}  (b) illustrates the entire region $\mathcal{P}_t\dfn \cup_{k=0}^{5} \mathcal{T}_k$ formed by $m_t+1=6$ triangular subsets that covers completely the region $\mathcal{A}$, i.e., $\mathcal{A}\subseteq \mathcal{P}_t$. 

%\[\mathcal{T}_0,\mathcal{T}_1,\mathcal{T}_2,\mathcal{T}_3,\mathcal{T}_4,\mathcal{T}_5\]
\subsubsection{Adaptive sampling}
\label{secAdS2}
To generate samples uniformly in $\mathcal{P}_t$, we first have to select a triangle proportionally to the areas $|\mathcal{T}_k|$, $k=0,...,m_t$. Therefore, we define the normalized weights
 \begin{equation}
 w_k\dfn \frac{|\mathcal{T}_k|}{\sum_{i=0}^{m_t}|\mathcal{T}_i|}, 
\end{equation}
and then we choose a triangular piece by drawing an index $k'\in\{0,...,m_t\}$ from the probability distribution $P(k)=w_k$. Using the method in the Appendix, we can easily generate a point $(v',u')$ uniformly in the selected triangular region $\mathcal{T}_{k'}$. If this point $(v',u')$ belongs to $\mathcal{A}$, we accept the sample $x'=v'/u'$ and set $m_{t+1}=m_{t}$, $\mathcal{S}_{t+1}=\mathcal{S}_{t}$ and $\mathcal{P}_t=\mathcal{P}_{t}$. Otherwise, we discard the sample $x'=v'/u'$ and incorporate it to the set of support points, $\mathcal{S}_{t+1}=\mathcal{S}_{t}\cup \{x'\}$, so that $m_{t+1}=m_{t}+1$ and the region $\mathcal{P}_t$ is improved by adding another triangle.

\subsubsection{Summary of the algorithm}
The adaptive RS algorithm to generate $N$ samples from $p(x)$ using RoU is summarized as follows.

\begin{enumerate}
\item \textbf{Initialization.} Start with $i=1$, $t=0$ and choose $m_1$ support points, $\mathcal{S}_{1}=\{s_{1},..., \ s_{m_1}\}$. 
\item \textbf{Iteration.} For $t \geq 1$, take the following steps.
	\begin{itemize}
	\item From $\mathcal{S}_t$, determine the intervals $\mathcal{I}_0, \ldots, \mathcal{I}_{m_t}$.
	\item Compute the upper bounds $L_k^{(j)}$, $j=1,2,3$, for each $k=0,...,m_t$ (see Section \ref{otherBound} and the Appendix). 
	\item Construct the triangular regions $\mathcal{T}_k$ as described in Section \ref{secConT}, $k=0,...,m_t$.
	\item Calculate the area $|\mathcal{T}_k|$ of every triangle, and compute the normalized weights
	\begin{equation}
        w_k\dfn \frac{|\mathcal{T}_k|}{\sum_{i=0}^{m_t}|\mathcal{T}_i|}
       \end{equation}
       with $k=0,...,m_t$.
       \item Draw an index $k'\in\{0,...,m_t\}$ from the probability distribution $P(k)=w_k$.
       \item Generate a point $(v',u')$ uniformly from the region $\mathcal{T}_{k'}$ (see the Appendix).
	\item  If $u'\leq \sqrt{p\left(\frac{v'}{u'}\right)}$, then accept the sample $x^{(i)}=x'=\frac{v'}{u'}$, set $i=i+1$, $\mathcal{S}_{t+1}=\mathcal{S}_{t}$ and $m_{t+1}=m_t$.
	\item Otherwise, if $u'> \sqrt{p\left(\frac{v'}{u'}\right)}$, then reject the sample $x'=\frac{v'}{u'}$, set $\mathcal{S}_{t+1}=\mathcal{S}_{t}\cup \{x'\}$ and $m_{t+1}=m_t+1$. 
	\item Sort $\mathcal{S}_{t+1}$ in ascending order and increment $t=t+1$. If $i>N$, then stop the iteration.
	\end{itemize}
\end{enumerate}

It is interesting to note that the region $\mathcal{P}_t$ is equivalent, in the domain of $x$, to a proposal function $\pi_t(x)$ formed by pieces of reciprocal uniform distributions scaled and translated (as seen in Section \ref{sRoU}), i.e., $\pi_t(x)\propto 1/(\lambda_k x+\beta_k)^2$ in every interval $\mathcal{I}_k$, for some $\lambda_k$ and $\beta_k$. 

%%%%%%%%%%%%%%%%%%%%%%%%%%%%
\subsection{Bounds for other potentials}
 \label{otherBound} 
%%%%%%%%%%%%%%%%%%%%%%%%%%%%
%How explained above and in Section \ref{secRS} in order to draw from the target density $p(x|\textbf{y})$ applying the ratio-of-uniforms method, in general we need to find bounds for the functions $\sqrt{p(x|\textbf{y})}$ and $x\sqrt{p(x|\textbf{y})}$. Specifically, since $p(x|\textbf{y})\geq 0$, we only need the upper bounds  $L^{(1)}\geq \sqrt{p(x|\textbf{y})}$, $L^{(2)}\geq x\sqrt{p(x|\textbf{y})}$  and $L^{(3)}\geq -x\sqrt{p(x|\textbf{y})}$ (note that $-L^{(3)}$ is a lower bound for $x\sqrt{p(x|\textbf{y})}$).
In this section we provide the details on the computation of the bounds $L_{k}^{(j)}$, $j=1,2,3$, needed for the implementation of the algorithm above.

We associate a potential $V^{(j)}$ to each function of interest. Specifically, since $p(x) \propto \exp\{-V(x;\textbf{g})\}$ we readily obtain that 
\begin{gather}
\begin{split}
\sqrt{p(x)}\propto \exp\big\{-V^{(1)}(x;\textbf{g})\big\}, \  \  &\mbox{with} \  \ V^{(1)}(x;\textbf{g})\dfn \frac{1}{2}V(x;\textbf{g}), \\
x\sqrt{p(x)}\propto \exp\big\{-V^{(2)}(x;\textbf{g})\big\}, \  \  &\mbox{with} \  \ V^{(2)}(x;\textbf{g})\dfn \frac{1}{2}V(x;\textbf{g})-\log(x), (x>0),\\
-x\sqrt{p(x)}\propto \exp\big\{-V^{(3)}(x;\textbf{g})\big\}, \  \  &\mbox{with} \  \ V^{(3)}(x;\textbf{g})\dfn \frac{1}{2}V(x;\textbf{g})-\log(-x), (x<0),\\
\end{split}
\end{gather}
respectively.

Note that it is equivalent to maximize the functions $\sqrt{p(x)}$, $x\sqrt{p(x)}$, $-x\sqrt{p(x)}$  w.r.t. $x$ and to minimize the corresponding potentials $V^{(j)}$, $j=1,2,3$, also w.r.t. $x$.  As a consequence, we may focus on the calculation of lower bounds $\gamma^{(j)}_k \leq V^{(j)}(x;\textbf{g})$ in an interval $x\in \mathcal{I}_k$, related to the upper bounds as $L^{(j)}_k=\exp\{-\gamma^{(j)}_k\}$, $j=1,2,3$ and $k=0,...,m_t$. This problem is far from trivial, though. Even for very simple marginal potentials, $\bar{V}_i$, $i=1,\ldots, n$, the potential functions, $V^{(j)}$, $j=1,2,3$, can be highly multimodal w.r.t. $x$ \cite{MartinoSigPro10}. 

In the Appendix we describe a procedure to find a lower bound $\gamma_k$ for the potential $V(x;\textbf{g})$.
We can apply the same technique to the function $V^{(1)}(x;\textbf{g})= \frac{1}{2}V(x;\textbf{g})$, associated to the function $\sqrt{p(x)}$, since $V^{(1)}$ is a scaled version of the generalized system potential $V(x;\textbf{g})$. Therefore, we can easily compute a lower bound $\gamma^{(1)}_{k}\leq V^{(1)}(x;\textbf{g})$ in the interval $\mathcal{I}_k$. 

The procedure in the Appendix can also be applied to find upper bounds for $x\sqrt{p(x)}$, with $x> 0$, and $-x\sqrt{p(x)}$ with $x<0$. Indeed, recalling that the associated potentials are
$V^{(2)}(x;\textbf{g})= \frac{1}{2}V(x;\textbf{g})-\log(x)$, $x>0$, and
$V^{(3)}(x;\textbf{g})= \frac{1}{2}V(x;\textbf{g})-\log(-x)$,
 $x< 0$, it is straightforward to realize that the corresponding modified potentials 
\begin{gather}
\begin{split}
&V^{(2)}(x;\textbf{r}_k)\dfn \frac{1}{2}V(x;\textbf{r}_k)-\log(x), \\
&V^{(3)}(x;\textbf{r}_k)\dfn \frac{1}{2}V(x;\textbf{r}_k)-\log(-x), 
\end{split}
\end{gather}
are convex in $\mathcal{I}_k$, since the functions $-\log(x)$  ($x>0$) and $-\log(-x)$  ($x<0$), are also convex. Therefore, it is  always possible to compute lower bounds $\gamma^{(2)}_{k}\leq V^{(2)}(x;\textbf{g})$ and  $\gamma^{(3)}_{k}\leq V^{(3)}(x;\textbf{g})$.
 
The corresponding upper bounds are $L^{(j)}_{k}=\exp\{-\gamma^{(j)}_k\}$, $j=1,2,3$ for all $x\in \mathcal{I}_k$. 
 
 \subsection{Heavier tails}
 \label{HeavyTalisSec}
 The standard RoU method can be easily generalized in this way: given the area 
 \begin{equation}
 \mathcal{A}_\rho= \Big\{ (v,u): 0\leq u \leq \left[p(v/u^\rho)\right]^{1/(\rho+1)}\Big\}, 
\end{equation}
if we are able to draw uniformly from it the pair $(v,u)$ then $x=\frac{v}{u^\rho}$ is a sample form $p_o(x)\propto p(x)$. 

The region $\mathcal{A}_\rho$ is bounded if $\left[p(x)\right]^{1/(\rho+1)}$ and $x\left[p(x)\right]^{\rho/(\rho+1)}$ are bounded. It occurs when $p(x)$ is bounded and the its tails decay as $1/x^{(\rho+1)/\rho}$  or faster. Hence, for $\rho > 1$ we can handle pdf's with
fatter tails than with the standard RoU method.

In this case the potentials associated to our target pdf $p_o(x)\propto \exp\{-V(x;\textbf{g})\}$ are
\begin{gather}
\begin{split}
&V^{(1)}(x;\textbf{g})\dfn \frac{1}{\rho+1}V(x;\textbf{g}), \\
&V^{(2)}(x;\textbf{g})\dfn \frac{\rho}{\rho+1}V(x;\textbf{g})-\log(x),  \mbox{ } \mbox{for} \mbox{ } x>0\\
&V^{(3)}(x;\textbf{g})\dfn \frac{\rho}{\rho+1}V(x;\textbf{g})-\log(-x), \mbox{ } \mbox{for} \mbox{ } x<0
\end{split}
\end{gather}
Obviouly, we can use the technique in the Appendix to obtain lower bounds and to build an extended adaptive RoU scheme to tackle target pdf's with fatter tails. Moreover, the constant parameter $\rho$ affects to the shape of $ \mathcal{A}_\rho$ and, as a consequence, also the acceptance rate of a rejection sampler. Hence, in some case it is interesting to find the optimal value of $\rho$ that maximizes the acceptance rate.

\section{Examples}
\label{sExample}
In this section we illustrate the application of the proposed techniques. The first example is devoted to compare the performance of the first adaptive technique in Section \ref{AltStrat} and the adaptive RoU algorithm in Section \ref{SecAdScheme} using an artificial model.  In the second example, we apply adaptive RoU scheme as a building block of an accept/reject particle filter \cite{Kunsch05} for inference in a financial volatility model. In this second example the first adaptive technique cannot implemented. Moreover, note that the other generalizations of the standard ARS method proposed in \cite{Evans98,Gilks92,Gorur08rev,Hoermann95} cannot be applied in both examples. 
%In the last example we study another stochastic volatility model where only the second  adaptive scheme can be applied. Therefore, we run an accept/reject a particle filter built around the proposed adaptive RoU algorithm. %Moreover, we remark that, in general, for different reasons the previous designed algorithms \cite{Gilks92,Evans98,Gorur08rev,MartinoSigPro10} can not applied.   

\subsection{Artificial example}
Let $x$ be a positive scalar signal of interest, $x\in\mathbb{R}^{+}$, with exponential prior, $q(x)\propto \exp\{-\lambda x\}$, $\lambda>0$, and consider the system with three observations, $\textbf{y}=[y_1,y_2,y_3]\in \mathbb{R}^{n=3}$, 
\begin{equation}
\label{sistejemplo}
y_1=a\exp(-bx)+\vartheta_1, \ \ y_2=c\log(dx+1)+\vartheta_2,\ \ \mbox{and}\mbox{ }\mbox{ } y_3=(x-e)^2+\vartheta_3,
\end{equation}
where $\vartheta_1$, $\vartheta_2$, $\vartheta_3$ are independent noise variables and $a$, $b$, $c$, $d$, $e$ are constant parameters.
Specifically, $\vartheta_1$ and  $\vartheta_2$ are independent with generalized gamma pdf $\Gamma_g(\vartheta_i;\alpha_i,\beta_i)\propto\vartheta_i^{\alpha_i}\exp\{-\vartheta_i^{\beta_i}\}$, $i=1,2$, with parameters $\alpha_1=4$, $\beta_1=2$ and $\alpha_2=2$, $\beta_2=2$, respectively. The variable $\vartheta_3$ has a Gaussian density $N(\vartheta_3;0,1/2)\propto\exp\{-\vartheta_3^2\}$.  

Our goal is to generate samples from the posterior pdf $p(x|{\bf y})\propto p({\bf y}|x)q(x)$. Given the system (\ref{sistejemplo}) our target density can be written as
\begin{gather}
\begin{split}
p_o(x)=p(x|{\bf y})\propto \exp\{-V(x;\textbf{g})\}, 
\end{split}
\end{gather}
where the potential is
\begin{gather}
\begin{split}
\label{potej1}
V(x;\textbf{g})=&(y_1-a\exp(-bx))^2-\log[y_1-a\exp(-bx))^4]+  \\
& +(y_2-c\log(dx+1))^2-\log[(y_2-c\log(dx+1))^2]+\\
& +(y_3-(x-e)^2)^2+\lambda x. \\
\end{split}
\end{gather}
We can interpreted it as 
\begin{equation}
V(x;\textbf{g})=\bar{V}_1(g_1(x))+\bar{V}_2(g_2(x))+\bar{V}_3(g_3(x))+\bar{V}_4(g_4(x))
\end{equation} 
where 
the marginal potentials are $\tilde{V}_1(\vartheta)=\vartheta^2-\log[\vartheta^4]$ with minimum at $\mu_1=\sqrt{2}$, $\tilde{V}_2(\vartheta)=\vartheta^2-\log[\vartheta^2]$ with minimum at $\mu_2=1$, $\tilde{V}_3(\vartheta)=\vartheta^2$ with minimum at $\mu_3=0$ and $\tilde{V}_4(\vartheta)=\lambda|\vartheta|$ with minimum at $\mu_4=0$ (we recall that $\lambda>0$ and $x\in\mathbb{R}^+$).  Moreover, the vector of nonlinearities is
\begin{equation}
\textbf{g}(x)=[g_1(x)=a\exp(-bx),g_2(x)=c\log(dx+1),g_3(x)=(x-e)^2,g_4(x)=x].
\end{equation}
Since all marginal potentials are convex and all nonlinearities are convex or concave, we can apply the proposed adaptive RoU technique. Moreover, since $\exp\{-\bar{V}_4(g_4(x))\}=\exp\{-\lambda x\}$ is easy to draw from, we can also applied the first technique described in Section \ref{AltStrat}.  

%It is possible to apply the two proposed adaptive schemes to draw from the posterior pdf in Eq. (\ref{likeEq}) and, therefore, we use this example to compare them directly. 

Note, however, that 
\begin{enumerate}
\item the potential $V(x;{\bf g})$ in Eq. (\ref{potej1}) is not convex,
\item the target function  $p(x|{\bf y})\propto\exp\{-V(x;{\bf g})\}$ can be multimodal,
\item the tails of the potential $V(x;{\bf g})$ are not convex and
%\item the potential $V(x;{\bf g})$ cannot express as sum of concave and convex functions and
\item it is not possible to study analytically the first and the second derivatives of the potential $V(x;{\bf g})$.
\end{enumerate}
Therefore, the techniques in \cite{Gilks92,Evans98,Gorur08rev,Hoermann95} and the basic strategy in \cite{MartinoStatCo10} cannot be used. %But, the alternative approach in Section \ref{AltStrat} proposed also in \cite{MartinoStatCo10} and the adaptive RoU technique can be applied.

The procedure in Section \ref{AltStrat} can be used because of the pdf $q_4(x)\propto \exp\{-V_4(g_4(x))\}=\exp\{-x\}$ can be easily integrated and sampled, even if it is restricted in a interval $\mathcal{I}_k$. In this case, the reduced potential is $V_{-4}(x;{\bf g})=\bar{V}_1(g_1(x))+\bar{V}_2(g_2(x))+\bar{V}_3(g_3(x))$  and the corresponding function $\exp\{-V_{-4}(x;{\bf g})\}$ is the likelihood, i.e., $p({\bf y}|x)=\exp\{-V_{-4}(x;{\bf g})\}$, since $q_4(x)$ coincides with the prior pdf. 
Moreover, in order to use the RoU method we additionally need to study the potential functions $V^{(1)}(x;\textbf{g})=\frac{1}{2}V(x;\textbf{g})$ and $V^{(2)}(x;\textbf{g})=\frac{1}{2}V(x;\textbf{g})-\log(x)$.  Since we assume $x\geq 0$, it is not necessary to study $V^{(3)}$ as described in Section \ref{otherBound}. 

\begin{figure*}[htb]
\centerline{
%	\subfigure[]{\includegraphics[width=5.5cm]{DensityEjemplo.pdf}}	
	\subfigure[]{\includegraphics[width=5.5cm]{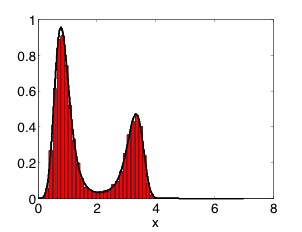}}		
%	\subfigure[]{\includegraphics[width=5.5cm]{ApproxlikeEjemplo.pdf}}	
	\subfigure[]{\includegraphics[width=5.5cm]{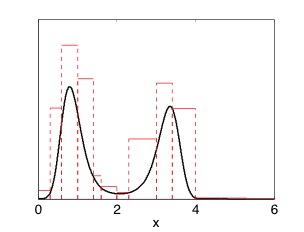}}	
%	\subfigure[]{\includegraphics[width=5.5cm]{TriangulosEjemplo.pdf}}	
	\subfigure[]{\includegraphics[width=5.5cm]{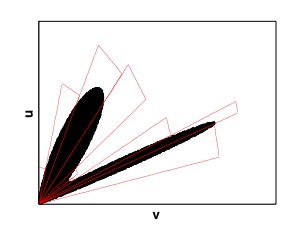}}		
}		
\caption{\textbf{(a)} The target function $p_o(x)=p(x|{\bf y})$ and the normalized histogram obtained by the adaptive RoU scheme. \textbf{(b)} The function $\exp\{-V_{-4}(x;{\bf g})\}$ obtained using the reduced potential $V_{-4}(x;{\bf g})=V(x;{\bf g})-\bar{V}_4(g_4(x))$ and its constant upper bounds $\exp\{-\gamma_k\}$, $k=0,...,m_t$. In this example, the  function $p({\bf y}|x)=\exp\{-V_{-4}(x;{\bf g})\}$ coincides with the likelihood. \textbf{(c)} The set $\mathcal{A}$ corresponding to the target pdf $p(x|{\bf y})$ using the RoU method and the region $\cup_{k=0}^{m_t=9} \mathcal{T}_k$, formed by triangles, constructed using the second adaptive rejection sampling scheme.
}  
\label{Ejemplofig}
\end{figure*}

We set, e.g., $a=-2$, $b=1.1$, $c=-0.8$, $d=1.5$, $e=2$, $\lambda=0.2$ and $\textbf{y}=[2.314,1.6,2]$,
and we start with the set of support points $\mathcal{S}_0=\{0,2-\sqrt{2},2,2+\sqrt{2}\}$ (details of the initialization step are explained in \cite{MartinoStatCo10}).  Figure \ref{Ejemplofig} (a) shows the  posterior density $p(x|\textbf{y})$ and the corresponding normalized histogram obtained by the first adaptive rejection sampling scheme. Figure \ref{Ejemplofig} (b) depicts, jointly, the likelihood function $p(\textbf{y}|x)=\exp\{-V_{-4}(x;{\bf g})\}$ and its stepwise approximation $\exp\{-\gamma_k\}$ $\forall x\in\mathcal{I}_k$, $k=0,...,m_t$. The computation of the lower bounds $\gamma_k\leq V_{-4}(x;{\bf g})$ has been explained in Appendix and Section \ref{AltStrat}. % Figure \ref{Ejemplofig} (c) depicts the set $\mathcal{A}$ corresponding to the posterior density $p(x|\textbf{y})$ and the region $\cup_{k=0}^{m_t} \mathcal{T}_k$ formed by triangles constructed as described in Section \ref{SecondAdaptivesect}.

Figure \ref{Ejemplofig} (c) depicts the set $\mathcal{A}$ (solid) obtained by way of the RoU method, corresponding to the posterior density $p(x|\textbf{y})$ and the region $\cup_{k=0}^{m_t} \mathcal{T}_k$ formed by triangular pieces, constructed as described in Section \ref{SecondAdaptivesect} with $m_t=9$ support points. 
  
\begin{figure*}[htb]
\centerline{
%\subfigure[]{\includegraphics[width=5.5cm]{TriangulosEjemplo.pdf}}			
	%\subfigure[]{\includegraphics[width=5.5cm]{acceptancerates.pdf}}
       \subfigure[]{\includegraphics[width=5.5cm]{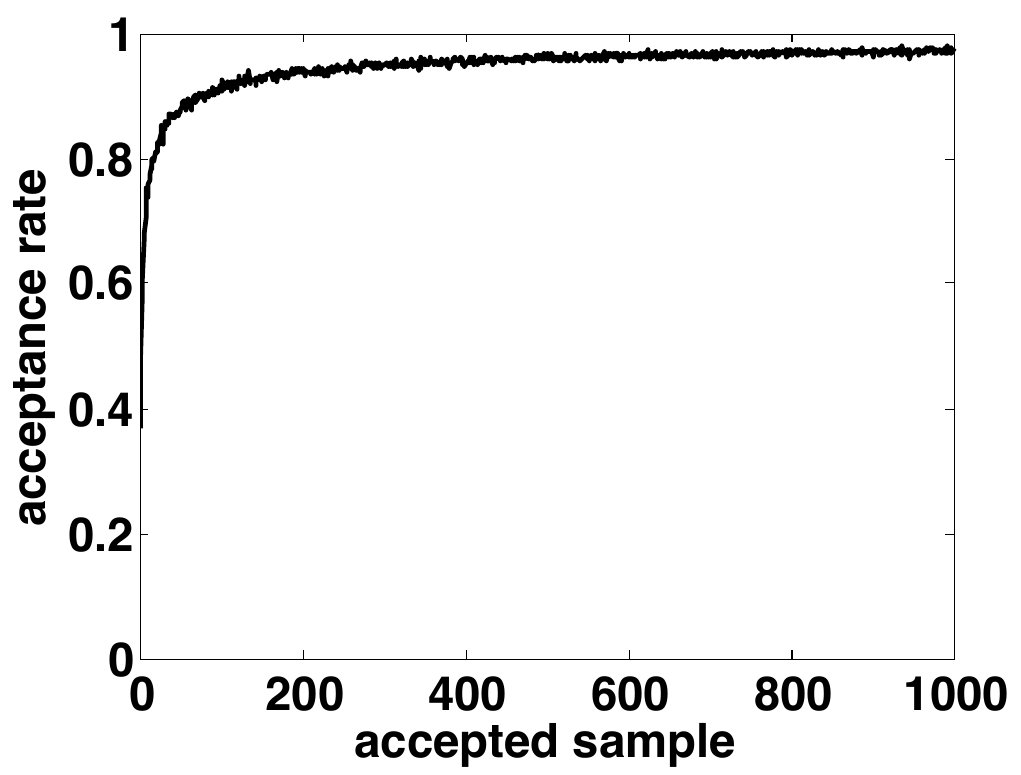}}	
       \subfigure[]{\includegraphics[width=5.5cm]{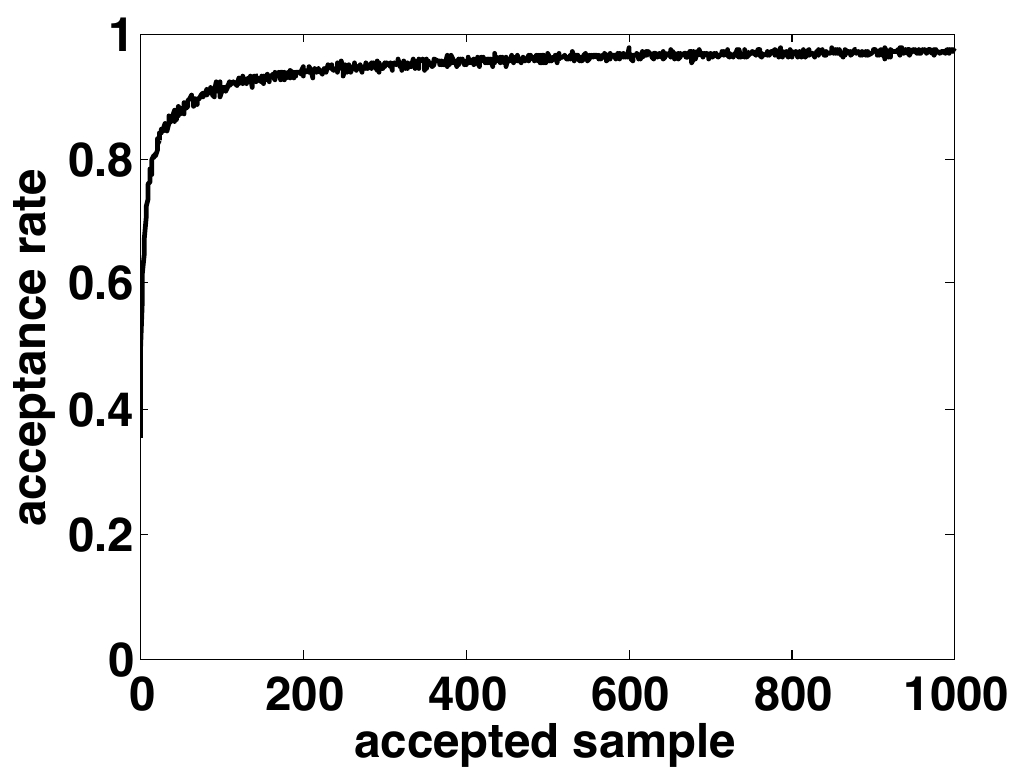}}	    
}		
\caption{\textbf{(a)} The curve of acceptance rates (averaged over 10,000 simulations) as a function of the first 1000 accepted samples using the first adaptive scheme described in Section \ref{AltStrat}. \textbf{(b)} The curve of acceptance rates (averaged over 10,000 simulations) as a function of the first 1000 accepted samples using the proposed adaptive RoU technique.}  
\label{Ejemplofig2}
\end{figure*}

The simulations shows that both method attain very similar acceptance rates\footnote{We define the acceptance rate $R_i$ as the mean probability of accepting the $i$-th sample from an adaptive RS algorithm. In order to estimate it, we have run $M=10,000$ independent simulations and recorded the numbers $\kappa_{i,j}$, $j=1,...,M$, of candidate samples that were needed in order to accept the $i$-th sample. Then, the resulting empirical acceptance rate is $\hat{R}_i=\sum_{j=1}^{M}\kappa_{i,j}^{-1} / M$.}.
%Since the proposed techniques improve the proposal pdf adding the rejected samples into the set of support points $\mathcal{S}_t$, this probability increases with the number of the accepted samples.}. 
In Figure \ref{Ejemplofig2} (a)-(b) are shown the curves of acceptance rates (averaged over 10,000 independent simulation runs) versus the first $1000$ accepted samples using the technique described in Section \ref{AltStrat} and the adaptive RoU algorithm explained in Section \ref{SecAdScheme}, respectively. More specifically, every point in these curves represent the probability of accepting the $i$-th sample. We can see that the methods are equal efficient, and the rates converge quickly close to 1. %In table \ref{ARfirstEje} are reported some acceptance rates obtained with the two techniques, for different samples.
 
%\begin{table}[!hbt]
%\begin{center}
%\caption{Acceptance Rates.}
%\label{ARfirstEje}
%\begin{tabular}{|c|c|c|c|c|}
%\hline
%$i$-th sample & 1 & 20 & 100 & 500\\
%\hline 
%First proposed technique  & 1.31 & 0.50 & 0.24 & 0.19 \\
%\hline
%Second proposed technique  & 0.92 & 0.38 & 0.22 & 0.18 \\
%\hline
%\end{tabular}
%\end{center}
%\end{table}
%We can see the curves grow quickly close to 1 and the performance are very similar. 
%The acceptance rate curves are very close therefore, for this reason, we illustrates in Figure \ref{Ejemplofig2} (b) the acceptance rates (averaged over 20,000 independent simulation runs) versus the number of accepted samples only for the first adaptive RS algorithm. 

\subsection{Stochastic volatility model}
\label{StocVolEj2}
%In this example, we study a different stochastic volatility model where only the adaptive RoU scheme can be applied. Let be $x_k\in\mathbb{R}^{+}$ the 
In this example, we study a stochastic volatility model where only the adaptive RoU scheme can be applied. Let be $x_k\in\mathbb{R}^{+}$ the volatility of a financial time series at time $k$, and consider the following state space system \cite[Chapter 9]{Doucet01b}-\cite{Jacquier94}
 with scalar observation, $y_k\in \mathbb{R}$, 
\begin{equation}
\label{sistejemploVol1}
\left\{
\begin{array}{l}
\log(x_k^2)=\beta\log(x_{k-1}^2)+\vartheta_{2,k}, \\
y_k=\log(x_k^2)+\vartheta_{1,k}, \\
\end{array}
\right.
\end{equation}
where $k\in\mathbb{N}$ denotes discrete time, $\beta$ is a constant, %$y_k=\log(z_k)$ with $z_k$ is the value of the time series at time $k$, 
$\vartheta_{2,k}\sim N(\vartheta_{2,k};0,\sigma^2)$ is a Gaussian noise, i.e., $p(\vartheta_{2,k})\propto \exp\{-\vartheta_{2,k}^2/2\sigma^2\}$, while $\vartheta_{1,k}$ has a density $p(\vartheta_{1,k})\propto \exp\{\vartheta_{1,k}/2-\exp(\vartheta_{1,k})/2\}$ obtained from the transformation $\vartheta_{1,k}=\log[\vartheta_{0,k}^2]$ of a standard Gaussian variable $\vartheta_{0,k}\sim N(\vartheta_{0,k};0,1)$.

Given the system in Eq. (\ref{sistejemploVol1}), the transition (prior) pdf  
\begin{equation}
\label{priorUltimoEje}
p(x_k|x_{k-1})\propto \exp\left\{-\frac{\log(x_k^2)-\beta \log(x_{k-1}^2)}{2\sigma^2}\right\},
\end{equation}
and the likelihood function 
\begin{gather}
\begin{split}
\label{likeEje2}
p(y_k|x_{k})\propto& \exp\left\{\frac{y_k-\log(x_k^2)}{2}-\frac{\exp(y_k-\log(x_k^2))}{2}\right\}=�\\
&=\exp\left\{\frac{y_k-\log(x_k^2)}{2}-\frac{\exp(y_k)-1/x_k^2}{2}\right\},
\end{split}
\end{gather}

We can apply the proposed adaptive scheme to implement an particle filter.
Specifically, let $\{x_{k-1}^{(i)}\}_{i=1}^N$ be a collection of samples from $p(x_{k-1}|y_{1:k-1})$. We can approximate the predictive density as 
\begin{eqnarray}
p(x_k|y_{1:k-1}) &=& \int p(x_k|x_{k-1})p(x_{k-1}|y_{1:k-1})dx_{k-1} \nonumber\\
&\approx& \frac{1}{N}\sum_{i=1}^N p(x_k|x_{k-1}^{(i)})
\end{eqnarray}
and then the filtering pdf as \cite{Doucet01}
%then the filtering pdf becomes
\begin{equation}
\label{eqFilteringk1}
p(x_k|y_{1:k}) \propto p(y_k|x_k) \frac{1}{N}\sum_{i=1}^N p(x_k|x_{k-1}^{(i)}). 
\end{equation}
If we can draw exact samples $\{x_k^{(i)}\}_{i=1}^N$ from (\ref{eqFilteringk1}) using a RS scheme, then the integrals of measurable functions $I(f)=\int f(x_k) p(x_k|y_{1:k})dx_k$ w.r.t. to the filtering pdf can be approximated as $I(f)\approx I_N(f)=\frac{1}{N}\sum_{i=1}^N f(x_k^{(i)})$. 

To show it, we first recall that the problem of drawing from the pdf in Eq. (\ref{eqFilteringk1}) can be reduced to generate an index $j\in \{1,...,N\}$ with uniform probabilities and then draw a sample from the pdf
\begin{equation}
\label{targetpdfej2}
%x_{k}^{(i)}\sim p(x_{k}|x_{k-1}^{(j*)},y_k) \propto p(y_k|x_k) p(x_k|x_{k-1}^{(j^*)}).
x_k^{(i)}\sim p_{o,j}(x_k) \propto p(y_k|x_k) p(x_k|x_{k-1}^{(j)}).
\end{equation}
Note that, in order to take advantage of the adaptive nature of the proposed technique, one can draw first $N$ indices $j_1,...,j_r,...j_N$, with $j_r\in \{1,...,N\}$, and then we sample $N_r$ particles from the same proposal, $x_k^{(m)}\sim p_{o,j_r}(x_k)$, $m=1,...,N_r$, where $N_r$ is the number of times the index $r\in \{1,...,N\} $ has been drawn. Obviously, $N_1+N_2+...+N_N=N$.  

\begin{figure*}[htb]
\centerline{
%	\subfigure[]{\includegraphics[width=5.4cm]{Fig2Eje3.pdf}}
%	\subfigure[]{\includegraphics[width=5.5cm]{FigEje3.pdf}}
	\subfigure[]{\includegraphics[width=5.4cm]{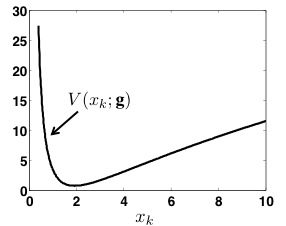}}
	\subfigure[]{\includegraphics[width=5.5cm]{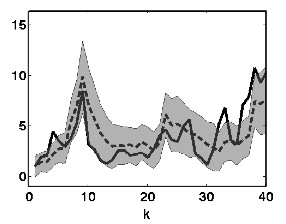}}	
      % \subfigure[]{\includegraphics[width=5.5cm]{acceptancerates2.pdf}}	
}		
\caption{\textbf{(a)} An example of potential function $V(x_k;\textbf{g})$ defined in Eq. (\ref{PotUltimoExample}), when $\alpha_k=1$, $y_k=2$ and $\sigma=0.8$. The right tail is concave. \textbf{(b)} An example of trajectory with $40$ time steps of the signal of interest $x_{k}$. The solid line illustrates the real trajectory generated by the system in Eq. (\ref{sistejemploVol1}) with $\beta=0.8$ and $\sigma=0.9$ while the dashed line shows the estimated trajectory attained by the particle filter with $N=1000$ particles. The shadowed area shows the standard deviation of the estimation.}  
\label{Ejemplofig4}
\end{figure*}

The potential function associated with the pdf in Eq. (\ref{targetpdfej2}) is 
\begin{gather}
\label{PotUltimoExample}
\begin{split}
V(x_k;\textbf{g})=-\frac{y_k-\log(x_k^2)}{2}+\frac{\exp\{y_k\}}{2}+\frac{1}{2x_k^2}+\frac{\big(\log(x_k^2)-\alpha_k \big)^2}{2\sigma^2} 
\end{split}
\end{gather}
where $\alpha_k=\beta\log\big[(x_{k-1}^{(j)})^2\big]$ is a constant and $\textbf{g}(x_k)=[g_1(x_k)\dfn \log(x_k^2),g_2(x_k)\dfn -\log(x_k^2)+\alpha_k]$. The potential function in Eq. (\ref{PotUltimoExample}) can be expressed as
\begin{equation}
V(x_k;\textbf{g})=\bar{V}_1\big(g_1(x_k)\big)+\bar{V}_2\big(g_2(x_k)\big),  
\end{equation}
where the marginal potentials are $\bar{V}_1(\vartheta_1)=\frac{1}{2}\big(-\vartheta_1+\exp\{\vartheta_1\}\big)$ and $\bar{V}_2(\vartheta_2)=\frac{1}{2\sigma^2}\vartheta_2^2$.
Note that in this example:
\begin{itemize}
\item Since the potential $V(x_k;\textbf{g})$ is in general a non-convex function and the study of the first and the second derivatives is not analytically tractable, the methods in \cite{Evans98,Gilks92} cannot be applied.
\item Moreover, since the potential function $V(x_k;\textbf{g})$ has concave tails, as shown in Figure \ref{Ejemplofig4} (a), the method in \cite{Gorur08rev} and the basic adaptive technique described in \cite{MartinoStatCo10} cannot be used either.
\item In addition, we also cannot apply the first adaptive procedure in Section \ref{AltStrat} because there are no simple (direct) techniques to draw from (and to integrate) the function $\exp\{-\bar{V}_1(g_1(x_k))\}$ or $\exp\{-\bar{V}_2(g_2(x_k))\}$.
\end{itemize}
 But, since the marginal potentials $\bar{V}_1$ and $\bar{V}_2$ are convex and we also know the concavity of the nonlinearities $g_1(x_k)$ and $g_2(x_k)$, we can apply the proposed adaptive RoU scheme. 
%Since the potential $V(x_k;\textbf{g})$ is in general a non-convex function and the study of the first and the second derivatives is not analytically tractable, the methods in \cite{Evans98,Gilks92} can not be applied. Moreover, since the potential function $V(x_k;\textbf{g})$ has concave tails, as shown in Figure \ref{Ejemplofig4} (a), the method in \cite{Gorur08rev} and the basic adaptive technique described in \cite{MartinoStatCo10} can not be used either. 
Therefore, the used procedure can be summarized in this way: at each time step $k$, we draw $N$ time from the set of index $r\in\{1,...,N\}$ with uniform probabilities and denote as $N_r$ the number of repetition of index $r$  ($N_1+...+N_N=N$). Then we apply the proposed adaptive RoU method, described in Section \ref{SecondAdaptivesect}, to draw $N_r$ particles from each pdf of the form of Eq. (\ref{targetpdfej2}) with $j=r$ and $r=1,....,N$.

Setting the constant parameters as $\beta=0.8$ and $\sigma=0.9$, we obtain an acceptance rate $\approx 42 \%$ (averaged over $40$ time steps in $10,000$ independent simulation runs). It is important to remark that it is not easy to implement a standard particle filter to make inference directly about $x_k$ (not about $\log(x_k)$), because it is not straightforward to draw from the prior density in Eq. (\ref{priorUltimoEje}). Indeed,  there are no direct methods to sample from this prior pdf and, in general, we need to use a rejection sampling or a MCMC approach. 

Figure \ref{Ejemplofig4} (b) depicts $40$ time steps of a real trajectory (solid line) of the signal of interest $x_{k}$ generated by the system in Eq. (\ref{sistejemploVol1}), with $\beta=0.8$ and $\sigma=0.9$. In dashed line, we see the estimated trajectory obtained by the particle filter using the adaptive RoU scheme with $N=1000$ particles. The shadowed area illustrates the standard deviation of the estimation obtained by our filter. The mean square error (MSE) achieved in the estimation of $x_k$  using $N=1000$ particles is $1.48$.

The designed particle filter is specially advantageous w.r.t. to the bootstrap filter when there is a significant discrepancy between the likelihood and prior functions.

\section{Conclusions}
\label{sConclusions}
We have proposed two adaptive rejection sampling schemes that can be used to draw exactly from a large family of pdf�s, not necessarily log-concave. Probability distributions of this family appear in many inference problems as, for example, localization in sensor networks \cite{Kunsch05,MartinoStatCo10,MartinoSigPro10}, stochastic volatility \cite[Chapter 9]{Doucet01b}, \cite{Jacquier94} 
or hierarchical models \cite[Chapter 9]{Gilks95bo}, \cite{Everson94,Fruhwirth09,Gamerman97}.

The new method yields a sequence of proposal pdfs that converge towards the target density and, as a consequence, can attain high acceptance rates. Moreover, it can also be applied when the tails of the target density are log-convex.  

The new  techniques are conceived to be used within more elaborate Monte Carlo methods. It enables, for instance, a systematic implementation of the accept/reject particle filter of \cite{Kunsch05}. There is another generic application in the implementation of the Gibbs sampler for systems in which the conditional densities are complicated. Since they can be applied to multimodal densities also when the tails are log-convex, these new techniques has an extended applicability compared to the related methods in literature \cite{Evans98,Gilks92,Gorur08rev,Leydold03,MartinoStatCo10}, as shown in the two numerical examples.
 
The first adaptive approach in Section \ref{AltStrat} is easier to implement. However, the proposed adaptive RoU technique is more general than the first scheme, as show in our application to a stochastic volatility model. Moreover using the adaptive RoU scheme to generate a candidate sample we only need to draw two uniform random variates but, in exchange, we have to find bound for three (similar) potential functions.

\section{Acknowledgements}
This work has been partially supported by the Ministry of Science and Innovation of Spain (project DEIPRO, ref. TEC-2009-14504-C02-01 and program Consolider-Ingenio 2010 CSD2008-00010 COMONSENS).
This work was partially presented at the Valencia 9 meeting.

\section{Appendix}
\subsection{Calculation of lower bounds}
\label{sectSystpot}

In some cases, it can be useful to find a lower bound $\gamma_k\in \mathbb{R}$ such that
\begin{equation}
\gamma_k\leq \min_{x\in \mathcal{I}_k} V(x;\textbf{g}),
\end{equation}
in some interval $\mathcal{I}_k=[s_{k},s_{k+1}]$.   

\begin{figure*}[htb]
\centerline{
%	\subfigure[]{\includegraphics[width=6.5cm]{GARSgeneral0.pdf}}	
%		\subfigure[]{\includegraphics[width=6.5cm]{GARSgeneral.pdf}	}
	\subfigure[]{\includegraphics[width=6.5cm]{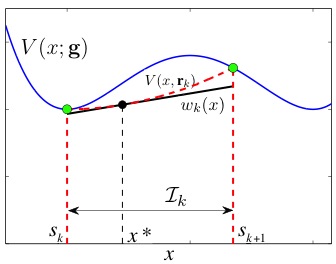}}	
		\subfigure[]{\includegraphics[width=6.5cm]{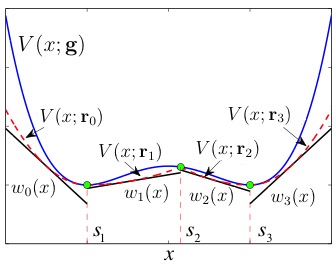}	}		
}
%\centerline{	
%	\includegraphics[width=7cm]{GARSgeneral.pdf}	
%}	
\caption{\textbf{(a)} Example of construction of the linear function $w_k(x)$ inside a generic interval $\mathcal{I}=[s_k,s_{k+1}]$. The picture shows a non-convex potential $V(x;\textbf{g})$ in solid line while the modified potential $V(x;\textbf{r}_k)$ is depicted in dashed line for $\forall x\in\mathcal{I}_k$. The linear function $w_k(x)$ is tangent to $V(x;\textbf{r}_k)$ at a arbitrary point $x^{*}\in \mathcal{I}_k$. \textbf{(b)} Example of construction of the piecewise linear function $W_{t}(x)$ with three support points $\mathcal{S}_t=\{s_1,s_2,s_{m_t=3}\}$. The modified potential $V(x;\textbf{r}_k)$, for $x\in \mathcal{I}_{k}$, is depicted with a dashed line. The piecewise linear function $W_{t}(x)$ consists of segments of linear functions $w_k(x)$ tangent to the modified potential $V(x;\textbf{r}_k)$.} 
\label{ARSfig}
\end{figure*}

\begin{figure*}[htb]
\centerline{
%	\subfigure[]{\includegraphics[width=6.5cm]{GARSgeneral_1.pdf}}
%	\subfigure[]{\includegraphics[width=6.95cm]{CotaConvex3.pdf}}
%		\includegraphics[width=6.95cm]{CotaConvex3.pdf}
		\includegraphics[width=6.95cm]{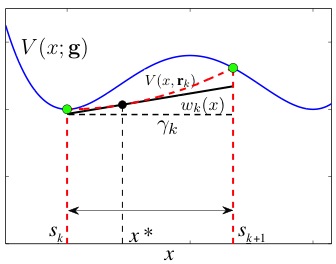}		
}
%\centerline{	
%	\includegraphics[width=7cm]{GARSgeneral.pdf}	
%}	
\caption{ %\textbf{(b)} 
The picture shows the potential $V(x;\textbf{g})$, the modified potential $V(x;\textbf{r}_k)$ an the tangent line $w_k(x)$ to the modified potential at an arbitrary point $x^{*}$ in $\mathcal{I}_{k}$. The lower bound $\gamma_{k}$ is obtained as $\gamma_{k}=\min[w_k(s_k),w_k(s_{k+1})]$. In this picture, we have $\gamma_k=w_k(s_k)$.
} 
\label{rectasfig}
\end{figure*}

If we are able to minimize analytically the modified potential $V(x;\textbf{r}_{k})$, we obtain
\begin{equation}
\gamma_{k}=\min_{x\in \mathcal{I}_{k}} V(x;\textbf{r}_{k})\leq \min_{x\in \mathcal{I}_{k}}V(x;\textbf{g}).
\end{equation}   
If the analytical minimization of the modified potential $V(x;\textbf{r}_{k})$ remains intractable, since the modified potential $V(x;\textbf{r}_k)$ is convex $\forall x\in \mathcal{I}_k$, we can use the tangent straight line $w_k(x)$ to $V(x;\textbf{r}_{k})$ at an arbitrary point $x^*\in \mathcal{I}_k$ to attain a bound. Indeed, a lower bound $\forall x\in \mathcal{I}_k=[s_{k},s_{k+1}]$ can be defined as
\begin{equation}
\gamma_{k}\dfn \min[w_k(s_k),w_k(s_{k+1})] \leq \min_{x\in \mathcal{I}_{k}} V(x;\textbf{r}_{k})\leq \min_{x\in \mathcal{I}_{k}}V(x;\textbf{g}).
\end{equation}
The two steps of the algorithm in Section \ref{GARSsect1} are described in the Figure \ref{ARSfig} (a) shows an example of construction of the linear function $w_k(x)$ in a generic interval $\mathcal{I}_k\subset \mathcal{D}$.
Figure \ref{ARSfig} (b) illustrates the construction of the piecewise linear function $W_{t}(x)$ using three support points, $m_t=3$.  Function $W_{t}(x)$ consists of segments of linear functions $w_k(x)$.
Figure \ref{rectasfig} (b) depicts this procedure to obtain a lower bound in an interval $\mathcal{I}_{k}$ for a system potential $V(x;\textbf{g})$ (solid line) using a tangent line (dotted line) to the modified potential $V(x;\textbf{r}_k)$ (dashed line) at an arbitrary point $x^{*}\in\mathcal{I}_{k}$.  
 
\subsection{Sampling uniformly in a triangular region}
\label{SampleTri}
Consider a triangular set $\mathcal{T}$ in the plane $\mathbb{R}^2$ defined by the vertices $\textbf{v}_1$, $\textbf{v}_2$ and $\textbf{v}_3$. % as shown in Figure \ref{figura2}. 
 We can draw uniformly from a triangular region \cite{Stein09}, \cite[p. 570]{Devroye86} with the following steps:
\begin{enumerate}
  \item Sample $u_1\sim \mathcal{U}([0,1])$ and  $u_2\sim \mathcal{U}([0,1])$.
  \item The resulting sample is generated by 
 \begin{equation}
  \textbf{x}'=\textbf{v}_1\min[u_1,u_2]+\textbf{v}_2(1-\max[u_1,u_2])+\textbf{v}_3(\max[u_1,u_2]-\min[u_1,u_2]) .
\end{equation}
\end{enumerate}
The samples $\textbf{x}'$ drawn with this convex linear combination are uniformly distributed within the triangle $\mathcal{T}$ with vertices $\textbf{v}_1$, $\textbf{v}_2$ and $\textbf{v}_3$.

\bibliography{bibliografia} 
   
\end{document}